\documentclass[useAMS,usenatbib,usegraphicx]{mn2e}
\usepackage{times}
\usepackage{epsfig}
\title[New age estimates of M31 globular clusters]
{New age estimates of M31 globular clusters from multi-colour
photometry}

\author[Z. Fan et al.]
{Z. Fan,$^{1,2}$ J. Ma,$^1$\thanks{E-mail:
majun@vega.bac.pku.edu.cn} R. de Grijs,$^3$
Y. Yang$^1$ and X. Zhou$^1$\\
$^1$National Astronomical Observatories, Chinese Academy of
Sciences, 20A Datun Road, Chaoyang District, Beijing 100012,
China\\
$^2$Graduate University of Chinese Academy of Sciences\\
$^3$Department of Physics \& Astronomy, The University of
Sheffield, Hicks Building, Hounsfield Road, Sheffield S3 7RH}

\date{Received; Accepted}

\pubyear{2006}

\begin{document}

\label{firstpage}

\maketitle

\begin{abstract}
The large majority of extragalactic star cluster studies performed to
date essentially use multi-colour photometry, combined with
theoretical stellar synthesis models, to derive ages, masses,
extinction estimates, and metallicities. M31 offers a unique
laboratory for studies of globular cluster (GC) systems. In this
paper, we obtain new age estimates for 91 M31 globular clusters, based
on improved photometric data, updated theoretical stellar synthesis
models and sophisticated new fitting methods. In particular, we used
photometric measurements from the Two Micron All Sky Survey (2MASS),
which, in combination with optical photometry, can partially break the
well-known age-metallicity degeneracy operating at ages in excess of a
few Gyr. We show robustly that previous age determinations based on
photometric data were affected significantly by this age-metallicity
degeneracy. Except for one cluster, the ages of our other sample GCs
are all older than 1 Gyr. Their age distribution shows populations of
young and intermediate-age GCs, peaking at $\sim 3$ and 8 Gyr
respectively, as well as the ``usual'' complement of well-known old
GCs, i.e., GCs of similar age as the majority of the Galactic GCs. Our
results also show that although there is significant scatter in
metallicity at any age, there is a noticeable lack of young metal-poor
and old metal-rich GCs, which might be indicative of an underlying
age-metallicity relationship among the M31 GC population.
\end{abstract}


\section{Introduction}
\label{Introduction.sec}

M31 is the nearest large spiral galaxy, at a distance of about 780 pc
\citep{sg98, mac01}. It contains more than 337 confirmed globular
clusters (GCs) and about 688 GC candidates \citep{gall04}, i.e.,
significantly more than in our own Galaxy. Thus, M31 provides an
excellent opportunity to study the properties of a large sample of
GCs. From the observational evidence collected thus far \citep[see,
e.g.,][]{rich05}, the M31 GCs and their Galactic counterparts reveal
some striking similarities \citep{ffp94,dj97,bhh02}. For example, both
GC systems seem to have similar mass-to-light ratios, structural
parameters, and velocity dispersion -- luminosity relations \citep[see
also][]{degrijs05}.  Studies of GCs in M31 can not only throw light on
lots of questions about the formation, evolution and properties of M31
itself, including its mass, dynamics and chemical composition, they
can also improve our understanding of the formation and structure of
galaxies in general \citep{batt80}. In addition, GCs can provide us
with good samples of Population II stars characterised by homogeneous
abundances and histories, and with unique stellar dynamical conditions
for our study \citep{barmby01}. Therefore, GCs are so important that
they are considered as the fossils of the earliest stages of galaxy
formation \citep{bh00}. However, at the distance of M31, construction
of colour-magnitude diagrams (CMDs) below the main sequence turn-off,
the most reliable method for age determinations of stellar
populations, is extremely challenging for current state-of-the-art
instrumentation.  Here, one suffers from the dual effects of crowding
and the intrinsic faintness of the cluster stars
\citep{rich96,rich05,stephens01,bb04}, although we note that
\citet{brown04} presented a CMD down to the turn-off for a $\sim 10$
Gyr-old M31 GC.

Since the pioneering work of \citet{Tinsley68,Tinsley72} and
\citet{ssb73}, evolutionary population synthesis modeling has become a
powerful tool to interpret integrated spectrophotometric observations
of galaxies as well as their components \citep[see,
e.g.,][]{Anders04}. Comprehensive compilations of relevant current
model sets, such as e.g.  developed by \citet[][henceforth BC93,
BC96]{bc93,bc96}, \citet{Leitherer95}, and \citet{frv97}, were
provided by \citet{lei96} and \citet{ken98}. The evolution of star
clusters is usually modeled based on the ``simple stellar population''
(SSP) approximation\footnote{An SSP is defined as a single generation
of coeval stars formed from the same progenitor molecular cloud (thus
implying a single metallicity), and governed by a given initial mass
function (IMF).}, which is a highly robust approximation for old GCs
in particular \cite[e.g.][]{bh01}. 

\citet{Ma01,Ma02a,Ma02b,Ma02c} and \citet{jiang03} estimated the ages
of, respectively, 180 star clusters in M33 and 172 GC candidates in
M31 by comparing the SSP synthesis models of BC96 with the clusters'
integrated photometric measurements in the
Beijing-Arizona-Taiwan-Connecticut (BATC) photometric system.
\citet{Ma06a} also determined the ages and metallicities of 33 M31 GCs
and candidates using the updated method and updated SSP synthesis
models of \citet[henceforth BC03]{bc03}, and \citet{Ma06b} estimated
the age and reddening value of the M31 GC 037-B327 based on
photometric measurements in a large number of broad-band passbands
from the optical to the near-infrared.

From an observational point of view, the study of M31 GCs is
complicated, since in most cases we only have access to their
integrated spectra and photometry, and cannot study the resolved
stellar population. Therefore, we can only obtain the key physical
parameters, such as the age and metallicity, by analysis of the
integrated spectra or photometry. However, a large body of evidence
suggests that there is a strong age-metallicity degeneracy if only
optical photometry is used
\citep{ar96,wor94,kaviraj06}. A very useful method to break this
degeneracy is through the application of particular spectral
diagnostics based on the occurrence of individual stellar
absorption-line features
\citep[e.g.,][]{fab73,rose84,rose85,dtt89,worth94,jw95,va99,bc03}.

At the same time, observational GC spectral energy distributions
(SEDs) are affected by reddening,
an effect that is also difficult to separate from the combined effects
of age and metallicity \citep{calz97,vazde97,orig99}. However, if the
metallicity and reddening are derived accurately (and, ideally,
independently), these degeneracies are largely (if not entirely)
reduced, and ages can then also be estimated accurately based on a
comparison of multi-colour photometry spanning a significant
wavelength range \citep{degrijs03b,Anders04} with theoretical stellar
population synthesis models.

In this paper, we present new age estimates for 91 GCs of the
\citet{jiang03} sample, based on their 13 intermediate-band photometry
in the BATC system, combined with additional broad-band optical and
2MASS near-infrared photometry, and on the updated SSP synthesis
models of BC03. Section \ref{Photometry.sec} describes the
intermediate-band, broad-band and 2MASS photometry of our GC
sample. In Section \ref{Metal&Red.sec}, we describe the metallicity
and reddening data of the sample GCs; Section \ref{age.sec} includes a
description of the SSP models used, and of our method to estimate the
ages of the sample GCs. The main results and a discussion are also
presented in this section. We summarise and conclude the paper in
Section \ref{Conclusions.sec}.

\section{Intermediate-band, Broad-band and 2MASS Photometry of our GC
Sample}
\label{Photometry.sec}

\subsection{Selection of the GC Sample}

The GC sample used in this paper was selected from that of
\citet{jiang03}, who published their intermediate-band photometry in
13 passbands taken from the BATC Multi-colour Survey of the Sky. They
estimated the ages using BC96 for 172 M31 GCs, selected from the
Bologna catalogue \citep{batt87}. This catalogue, which includes a
total of 827 GC candidates, is the most comprehensive list of M31
GCs. Of these objects, 353 are considered class A (254) and class B
(99) objects, i.e., probable GCs (or candidates) to a very high and
high level of confidence, respectively; 152 sources are listed as
class C, and the others fall into the lower-confidence classes D and
E. \citet{jiang03} took all candidates in classes A and B as their
original GC sample. However, only 223 objects overlapped in position
with the BATC $58^{\prime}$ $\times $ $58^{\prime}$ CCD
field\footnote{The BATC field of view covers most of the visible disc
of M31; the consequence of the limited field of view for the present
paper is that we may have missed a number of GCs in the extreme outer
disc regions along the galaxy's major axis, and that the fraction of
outer halo GCs is underrepresented (see aslo Fig. 1).}. Four of these
are most likely stars rather than GCs \citep{bh00}. An additional 47
GCs were saturated in some of the BATC filters, and were therefore not
included in the final Jiang et al. sample. As a consequence, there are
172 GCs in their final sample \citep[see details from][]{jiang03}. For
the purpose of estimating accurate GC ages, we selected GCs from the
\citet{jiang03} sample for which the metallicity and reddening values
had been estimated accurately and homogeneously in previous studies.
\citet{bh00} and \citet{per02} derived metallicities of,
respectively 61 and more than 200 GCs and GC candidates in M31.
\citet[][P. Barmby, priv. comm.]{bh00} determined the reddening values
for 223 GC candidates. Based on cross-identification of the
\citet{jiang03} sample and that of \citet{bh00} and \citet{per02}, we
selected 93 objects as our initial M31 GC cluster sample. We refer the
reader to Section 3 for more technical details. When we compared the
multi-colour photometry of our sample GCs with theoretical stellar
synthesis models, we could not obtain a reasonable fit to the SED of
one GC, Bo131. A detailed investigation of the M31 images of BATC
Multi-colour Survey of the Sky reveals that this GC is superimposed
onto a very high and variable background, so that the resulting
photometry is highly uncertain. Therefore, we discarded this object
from our final sample. In addition, 037-B327 has been studied in
detail by \citet{Ma06b}, and is therefore also excluded from the
present sample. Our final sample in this paper thus contains 91 GCs.

\begin{figure}
\resizebox{\hsize}{!}{\rotatebox{-90}{\includegraphics{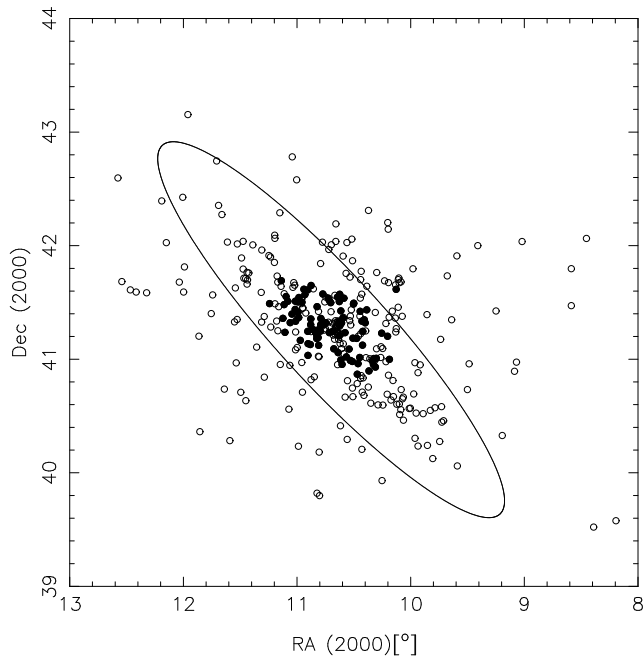}}}
\caption{Map of all known {\it bona fide} M31 GCs from \citet{gall04}
and the subsample of GCs discussed in this paper. Open and solid
circles present the \citet{gall04} GCs and our subsample,
respectively; the ellipse traces the $D_{25}$ isophote of the galaxy.}
\label{fig1}
\end{figure}

Since \citet{jiang03}'s GC sample selection was only limited by the
size of their field of view, their GC sample constitutes a random
selection of M31 GCs, provided that there are no significant
positional biases present in the M31 GC sample as a whole between the
disc and the outer halo. In order to illustrate this limitation,
Fig. 1 shows the map of all known {\it bona fide} M31 GCs from
\citet{gall04} and the subsample of GCs discussed in this paper. It is
clear that our sample GCs are indeed limited to the central region of
M31, as determined solely by the BATC observational field of view, and
without further selection criteria imposed that might jeopardise the
random nature of our sample selection. However, we point out that our
sample may be affected by an underrepresentation of young ($\la 1-2$
Gyr-old) blue luminous compact clusters (BLCCs). \citet{fp05} found
that such BLCCs seem to avoid the inner regions of the M31 disc, and
appear to be clearly projected onto the spiral-arm structure in the
outer disc (predominantly along the major-axis direction; see their
Fig. 5), with which they also share kinematic properties \citep[see
also][]{puz05}. In addition, there is evidence for a roughly 25 to 50
per cent solar-metallicity ([Fe/H] $= -0.6$ to $-0.2$ dex) population
of older GCs in the M31 halo \citep[e.g.][and references
therein]{puz05}, while the more metal-poor clusters appear to be
scattered throughout the entire disc-halo system
\citep[e.g.][]{bh00,per02,puz05}.

\begin{figure}
\resizebox{\hsize}{!}{\rotatebox{-90}{\includegraphics{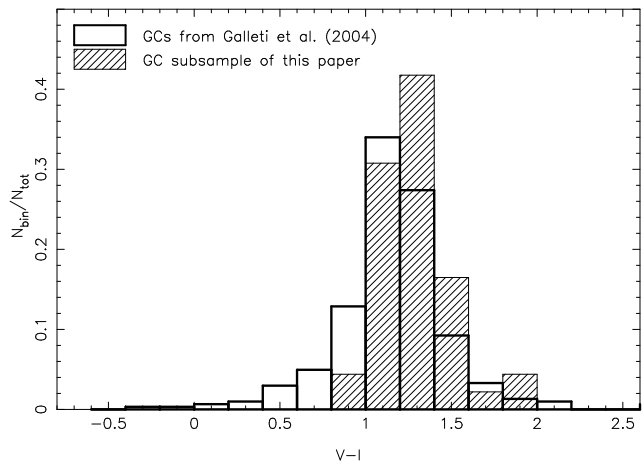}}}
\caption{$(V-I)$ distribution of all known {\it bona fide} M31 GCs
from \citet{gall04} and the subsample of GCs discussed in this paper.}
\label{fig2}
\end{figure}

Fig. \ref{fig2} compares the $(V-I)$ colour distributions of the
\citet{gall04} M31 GCs with that of our subsample. In order to show
any differences as clearly as possible, we use the normalised number
of GCs in each sample. Fig. \ref{fig2} shows that the $(V-I)$ colour
distribution of our subsample of M31 GCs is similar to that of the
full \citet{gall04} sample. If anything, the disc GCs, by which our
subsample is dominated, may be somewhat redder than the mean of the
entire M31 GC sample. This is not surprising in view of the possible
selection biases discussed above, while the main galactic disc of M31
may also give rise to redder GC colours owing to the presence of
fairly significant dust lanes (particularly closer to the galactic
centre).

In Section \ref{agecomp.sec}, we will further consider to which extent
our GC sample represents that of \citet{jiang03}. We also note that
although most present catalogues of M31 GCs are not complete,
\citet{fusipecci} have shown that the M31 GC population (including the
sample in the Bologna catalogue on which the present sample is based)
is currently fairly complete down to $V = 18$ mag ($M_V \sim -6.5$
mag). As such, we believe that we have not introduced additional
photometric biases into our sample selection by using the Bologna
catalogue as our master list of M31 GCs.

\subsection{Intermediate-band Photometry of Our Sample GCs}

The large-field multi-colour observations of M31 were obtained in the
BATC photometric system, obtained with the $60/90$cm f/3 Schmidt
telescope at the Xinglong Station of the National Astronomical
Observatories of China. A Ford Aerospace 2048$\times$2048 CCD camera,
with a 15 $\mu$m pixel size is mounted at the Schmidt focus. This
set-up provides a CCD field of view of $58^{\prime}$ $\times $
$58^{\prime}$, with a pixel size of
$1\arcsec{\mbox{}\hspace{-0.15cm}.} 7$. The filters were designed
specifically to avoid contamination from the brightest and most
variable night sky emission lines. \citet{jiang03} extracted 123
images of M31 from the BATC survey archive, taken in 13 BATC filters
at wavelengths from 4,000 to 10,000{\AA} between 1995 September and
1999 December. They combined multiple images through the same filter
to improve the image quality, and determined the magnitudes of 172 GCs
in the 13 BATC filters based on the combined images using standard
aperture photometry, i.e., in essence by employing the PHOT routine in
{\sc daophot} \citep{stet87}. The BATC photometric system calibrates
the zero-magnitude level in a similar fashion to the
spectrophotometric AB magnitude system. For flux calibration, the
Oke-Gunn primary flux standard stars HD 19445, HD 84937, BD
+26$^{\circ}$2606, and BD +17$^{\circ}$4708 \citep{ok83} were observed
on photometric nights \citep{yan00}. Table 2 of \citet{jiang03} lists
the resulting BATC photometry of their 172 GCs, including the 91 GCs
in this paper.

\subsection{Broad-band Photometry of Our Sample GCs}

In order to estimate the ages of our sample GCs accurately, we try to
use as many photometric data points covering as large a wavelength
range as possible \citep[cf.][]{degrijs03b,Anders04}. Using the
4-Shooter CCD mosaic camera and the SAO infrared imager on the 1.2m
telescope at the Fred Lawrence Whipple Observatory, \citet{bh00}
presented optical and infrared photometric data for 285 M31 GCs
\citep[see Table 3 of][]{bh00}. For most of our sample GCs,
photometric measurements in the $UBVRI$ bands were published by
\citet{bh00}. For the remaining GCs, we refer to \citet{gall04}, who
updated the Bologna Catalogue with the homogenised optical ($UBVRI$)
photometry collected from the most recent photometric references
available in the literature.
In this revised Bologna Catalogue of M31 globular clusters,
\citet{gall04} did not include the photometric
uncertainties. Therefore, we refer to the original works as indicated
in \citet{gall04}. For a number of GCs, for which $U$-band photometric
uncertainties were not presented in the original references, a
photometric uncertainty of 0.08 mag was adopted
\citep[see][]{gall04}. \citet{gall04} took the photometry in the
$UBVRI$ bands from \citet{bh00} as a reference to obtain a Master
Catalogue of photometric measurements compiled in as homogeneously a
fashion as possible. All other catalogues containing $UBVRI$
photometry were transformed to this reference by applying the offsets
derived from objects in common between the relevant catalogue and the
list of \citet{bh00}. Therefore, 
the measurements are internally consistent.

\subsection{2MASS Photometry of Our Sample GCs}

As pointed out by \citet{wor94}, the age-metallicity degeneracy in
optical broad-band colours is $\rm{\Delta age/\Delta Z\sim 3/2}$.
This implies that the composite spectrum of an old stellar population
is indistinguishable from that of a younger but more metal-rich
population (and vice versa) \citep[see
also][]{MacArthur04}. \citet{jong96} and \citet{Anders04} showed that
this degeneracy can be partially broken by adding infrared photometry
to the optical colours, depending on the age of the stellar
population. \citet{Cardiel03} found that the inclusion of an infrared
band can improve the predictive power of the stellar population
diagnostics by $\sim 30$ times compared to using optical photometry
alone. \citet{wu05} also showed that the use of near-infrared colours
can greatly contribute to break the age-metallicity degeneracy. In
this paper we add 2MASS photometry to our broad and intermediate-band
optical photometry to estimate the GC ages to the highest possible
accuracy.

Using the 2MASS database, \citet{gall04} identified 693 known and
candidate GCs in M31, and listed their 2MASS $JHK$s magnitudes.
\citet{gall04} transformed all 2MASS magnitudes to the CIT photometric
system \citep{Elias82,Elias83} using the colour transformations in
\citet{Carpenter01}. However, we need the original 2MASS $JHK$s
magnitudes for our sample GCs in order to compare our observational
SED to the SSP models, so we reversed this transformation using the
same procedures. Since \citet{gall04} did not give the photometric
2MASS $JHK$s uncertainties, we obtained photometric uncertainties by
comparing the photometric magnitudes with Fig. 2 of
\citet{Carpenteretal01}, who plot the observed photometric rms
uncertainties in the time series as a function of magnitude for stars
brighter than their observational completeness limits. In fact, the
photometric uncertainties adopted do not affect our results
significantly, as we will show below (see section 4.3 for details).

Finally, we point out that because of the very long wavelength
coverage from $U$ to the near-infrared, and the large number of
photometric data points for each GC, our age estimates, and in
particular the associated uncertainties are highly robust. In
fact, depending on the (internal) extinction curve adopted, they are
competitive with respect to spectroscopic age determinations based on
either a short wavelength range or a limited number of age diagnostics
(where small-scale differences in the continuum level may have
significant consequences for the age determinations), as discussed
extensively in \citet{degrijs03b} and \citet{dg05a}. \citet{ssb04}
provide convincing support for this statement by showing that
spectroscopic age determinations are not necessarily better or more
accurate than ages obtained from photometry.

\section{Metallicities and Reddening}
\label{Metal&Red.sec}

To estimate the ages of our sample GCs accurately, we required that
our GC sample have both independently determined spectroscopic
metallicities, and reddening values. We used two homogeneous reference
sources of spectroscopic metallicities, \citet{bh00} and
\citet{per02}.

\citet{bh00} present metallicities of 61 GC candidates, using the Keck
LRIS and the MMT Blue Channel spectrographs. With the Keck LRIS, they
used a 600 $\ell$ mm$^{-1}$ grating with a 1.2 {\AA} pixel$^{-1}$
dispersion from 3670--6200 {\AA}, and a resolution of 4-5 \AA. With
the MMT Blue Channel, they used a 300 $\ell$ mm$^{-1}$ grating with a
3.2 {\AA} pixel$^{-1}$ dispersion from 3400--7200 {\AA}, and a
resolution of 9-11 {\AA}. \citet{bh00} computed the absorption-line
indices following \citet{bh90}; subsequently, they used the
\citet{bh90} spectral index -- metallicity calibration. \citet{bh90}'s
metallicity calibration is based on six absorption-line indices
measured from integrated cluster spectra, which provides an [Fe/H]
estimate accurate to about 15 per cent.

\citet{per02} list metallicities for more than 200 GCs in M31 using
the Wide Field Fibre Optic Spectrograph at the 4.2 m William Herschel
Telescope, which provides a total spectral coverage of $\sim$
3700--5600 {\AA} with two gratings. One grating (H2400B, 2400 $\ell$
mm$^{-1}$) yields a dispersion of 0.8 {\AA} pixel$^{-1}$ and a
spectral resolution of 2.5 {\AA} over the range 3700--4500 {\AA}, and
the other grating (R1200R, 1200 $\ell$ mm$^{-1}$) is characterised by
a dispersion of 1.5 {\AA} pixel$^{-1}$ and a spectral resolution of
5.1 {\AA} over the range 4400--5600 {\AA}. \citet{per02} then
calculated 12 absorption-line indices following \citet{bh90}. By
comparison of the line indices with published M31 GC [Fe/H] values
from previous studies \citep{bonoli87,bh90,bh00}, they applied a
linear least-squares minimisation. Final cluster metallicities were
determined from an unweighted mean of the [Fe/H] values calculated
from the CH (G band), Mg$b$, and Fe53 line strengths.

We draw the reader's attention to the fact that the M31 GC
metallicities thus obtained were based on the [Fe/H] calibration of
\citet{bh90}, who included only old GCs in their calibration
sample. Since in this paper we estimate the ages of our M31 GC sample,
and find that some of them are young or of intermediate age, our age
determination may be somewhat biased. However, we point out that (i)
the main features in the M31 GC age distribution that we will derive
in Section(s) 4.5 (and 4.6) are similar to those derived by authors
who do not rely on the \citet{bh90} metallicity calibration
\citep[e.g.][]{bb05} -- on a one-to-one basis, the ages of most of the
GCs we have in common with these authors do not differ by more than
$1-2\sigma$, and (ii) we do not have the spectroscopic data required
to quantify this effect in detail; instead, we refer to \citet{bb05}
for an in-depth discussion on this issue.

For the reddening values of the sample GCs we refer to \citet{bh00},
who determined the reddening for each individual cluster using
correlations between optical and infrared colours and metallicity and
by defining various ``reddening-free'' parameters using their large
database of multi-colour photometry. \citet{bh00} found that the M31
and Galactic GC extinction laws, and the M31 and Galactic GC
colour-metallicity relations are similar to each other. They then
estimated the reddening to M31 objects with spectroscopic data using
the relation between intrinsic optical colours and metallicity for
Galactic clusters. For objects without spectroscopic data, they used
the relationships between the reddening-free parameters and certain
intrinsic colours based on the Galactic GC data. \citet{bh00} compared
their results with those in the literature and confirmed that their
estimated reddening values are reasonable, and quantitatively
consistent with previous determinations for GCs across the entire M31
disc. In particular, \citet{bh00} showed that the distribution of
reddening values as a function of position appears reasonable in that
the objects with the smallest reddening are spread across the disk and
halo, while the objects with the largest reddening are concentrated in
the galactic disk. Therefore, we adopted the reddening values from
\citet[][also P. Barmby, priv. comm.]{bh00} for our GC sample.

\section{Age determinations of the sample GCs}
\label{age.sec}

\subsection{Stellar Populations and Synthetic Photometry}

In evolutionary synthesis models, SSPs are modeled by a collection of
evolutionary tracks of stars with different initial masses and
chemical compositions, and a set of stellar spectra at different
evolutionary stages. To estimate the ages of our sample GCs we compare
their SEDs with the updated SSP models of BC03. BC03 provide the
evolution of the spectra and photometric properties of SSPs for a wide
range of stellar metallicities. The model set includes 26 SSP models
(both of high and low resolution) based on the 1994 Padova
evolutionary tracks, 13 of which were computed using the \citet{cha03}
IMF assuming lower and upper mass cut-offs of $m_{\rm L}=0.1$
M$_{\odot}$ and $m_{\rm U}=100$ M$_{\odot}$, respectively, while the
other 13 were computed using the \citet{sal55} IMF with the same mass
cut-offs. In addition, BC03 provide 26 SSP models using the 2000
Padova evolutionary tracks. However, as they point out, the 2000
Padova models tend to produce worse agreement with observed galaxy
colours. These SSP models contain 221 spectra describing the spectral
evolution of SSPs from $1 \times 10^5$ yr to 20 Gyr. The evolving
spectra include the contribution of the stellar component at
wavelengths from 91{\AA} to $160\mu$m. In this paper, we adopt the
high-resolution SSP models computed using the 1994 Padova evolutionary
tracks and a \citet{sal55} IMF\footnote{We note that because of the
slow SED evolution of SSPs at ages in excess of a few Gyr, all of the
most commonly used spectral synthesis models agree very well at these
ages. Therefore, the choice of IMF is {\it only} important for
estimating the photometric mass of the cluster, and does {\it not}
affect the determination of the ages of old GCs.}. We note that
although the current best estimate of the age of the Universe is of
order 13.7 Gyr, the SSP models and the stellar evolutionary tracks
that form their basis have been calculated for ages up to 20 Gyr. It
is not straightforward to correct for this discrepancy; one would need
to recalculate all stellar evolutionary tracks for all
metallicities. This situation is exacerbated by the fact that (i)
young open clusters, up to $\sim 5 $ Gyr, have been used to properly
and robustly constrain the stellar evolution at younger ages, and (ii)
at older ages the observational diagnostics are less sensitive to
changes in age, resulting in significantly greater uncertainties at
those ages. These are therefore issues that one needs to keep in mind
in the context of physical parameters derived for the older GCs.

Since our observational data are integrated luminosities through our
set of filters, we convolved the BC03 SSP SEDs with the BATC
intermediate-band, broad-band $UBVRI$ and 2MASS filter response curves
to obtain synthetic optical and near-infrared photometry for
comparison. The synthetic $i^{\rm th}$ filter magnitude can be
computed as
\begin{equation}
m_i=-2.5\log\frac{\int_{\lambda}F_{\lambda}\varphi_{i} (\lambda){\rm
d}\lambda}{\int_{\lambda}\varphi_{i}(\lambda){\rm d}\lambda}-48.60
\quad,
\end{equation}
where $F_{\lambda}$ is the theoretical SED and $\varphi_i$ the
response curve of the $i^{\rm th}$ filter of the BATC, $UBVRI$ and
2MASS photometric systems. Here, $F_{\lambda}$ varies with age and
metallicity.

\subsection{Fit Results}

We use a $\chi^2$ minimisation test to examine which BC03 SSP
models are most compatible with the observed SED, following
\begin{equation}
\chi^2=\sum_{i=1}^{21}{\frac{[m_{\lambda_i}^{\rm
obs}-m_{\lambda_i}^{\rm mod}(t)]^2}{\sigma_{i}^{2}}} \quad,
\end{equation}
where $m_{\lambda_i}^{\rm mod}(t)$ is the integrated magnitude in the
$i^{\rm th}$ filter of a theoretical SSP at age $t$,
$m_{\lambda_i}^{\rm obs}$ presents the observed integrated magnitude
in the same filter, and
\begin{equation}
\sigma_i^{2}=\sigma_{{\rm obs},i}^{2}+\sigma_{{\rm mod},i}^{2}
\quad .
\end{equation}
Here, $\sigma_{{\rm obs},i}^{2}$ is the observational uncertainty, and
$\sigma_{{\rm mod},i}^{2}$ is the uncertainty associated with the
model itself, for the $i^{\rm th}$ filter. \citet{charlot96} estimated
the uncertainty associated with the term $\sigma_{{\rm mod},i}^{2}$ by
comparing the colours obtained from different stellar evolutionary
tracks and spectral libraries. Following \citet{wu05} and
\citet{Ma06b}, we adopt $\sigma_{{\rm mod},i}^{2}=0.05$.

\begin{figure*}
\begin{center}
\includegraphics[angle=-90,width=160mm]{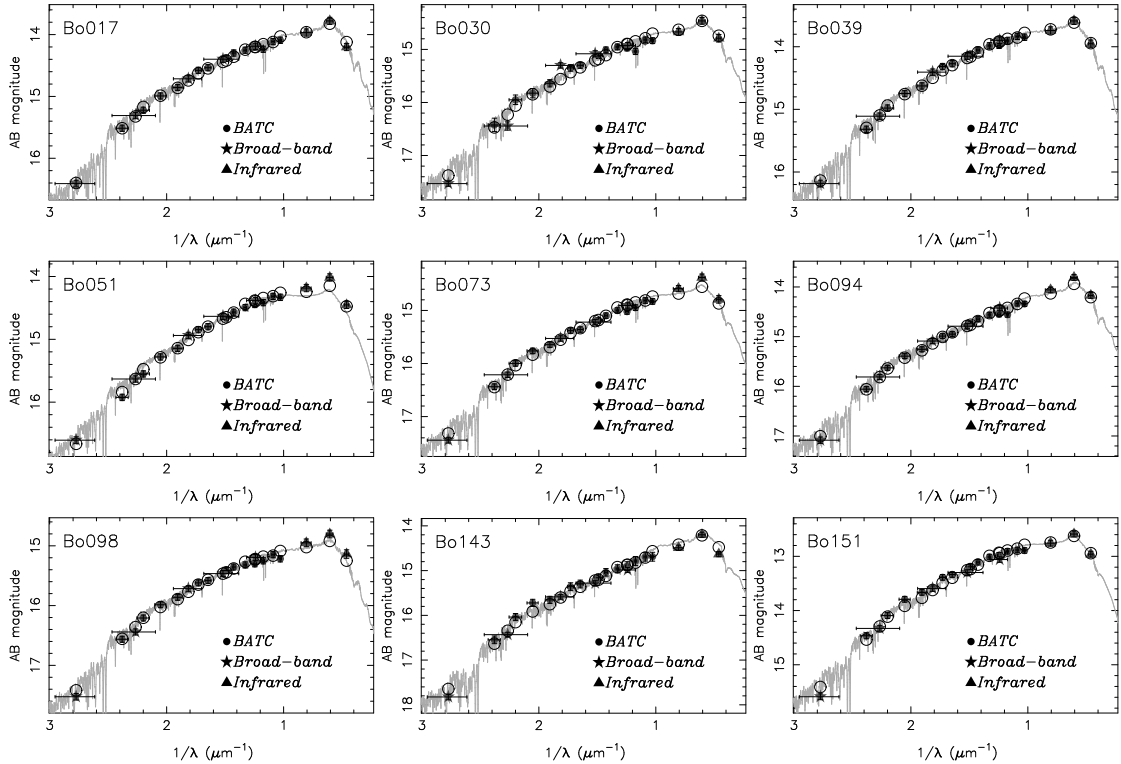}
\caption{Observational SEDs of a representative, randomly selected
subsample of M31 GCs, and the SEDs of the best-fitting models.}
\label{fig3}
\end{center}
\end{figure*}

The BC03 SSP models include six initial metallicities, 0.0001, 0.0004,
0.004, 0.008, 0.02 (= solar), and 0.05. Spectra for other
metallicities can be obtained by linear interpolation of the
appropriate spectra for any of these metallicities. For our sample
GCs, whose metallicity and reddening values were published by other
authors, the cluster age is the sole parameter to be estimated (for a
given IMF and extinction law, which we assume to be universal among
our GC sample and for the main calibrators). The values for the
extinction coefficient, $R_{\lambda}$, are obtained by interpolating
the interstellar extinction curve of \citet{car89}. In Fig. \ref{fig3}
we show the observational SEDs of a representative, randomly selected
subsample of M31 GCs, and the SEDs of the best-fitting models. The fit
results for our entire sample are listed in Tables 1 and 2.

\begin{table}
\caption{Fit Results for our 91 M31 GCs.} \label{t1.tab}
\begin{center}
\begin{tabular}{cccc}
\hline  ID&   Age &  Metallicity &   $\chi_{\rm min}^2$ \\
          & (Gyr) &    [Fe/H]    & (per degree of freedom) \\
\hline
Bo009 &   16.75$\pm$4.90&    -1.57$\pm$0.26&    2.87\\
Bo015 &    1.43$\pm$0.04&    -0.35$\pm$0.96&    1.86\\
Bo017 &    2.30$\pm$0.07&    -0.42$\pm$0.45&    0.59\\
Bo023 &   20.00$\pm$4.94&    -0.92$\pm$0.10&    2.69\\
Bo025 &    8.25$\pm$2.08&    -1.46$\pm$0.13&    1.62\\
Bo027 &   17.50$\pm$2.95&    -1.64$\pm$0.32&    3.80\\
Bo029 &   14.25$\pm$4.25&    -0.32$\pm$0.14&    5.51\\
Bo030 &    5.25$\pm$1.30&    -0.39$\pm$0.36&    3.81\\
Bo031 &    2.60$\pm$0.92&    -1.22$\pm$0.40&    4.56\\
Bo034 &   20.00$\pm$4.69&    -1.01$\pm$0.22&    3.46\\
Bo036 &   20.00$\pm$4.34&    -0.99$\pm$0.25&    2.13\\
Bo038 &    6.00$\pm$0.82&    -1.66$\pm$0.44&    0.43\\
Bo039 &    4.75$\pm$0.47&    -0.70$\pm$0.32&    0.61\\
Bo041 &   19.50$\pm$5.03&    -1.22$\pm$0.23&    0.76\\
Bo042 &    2.60$\pm$0.14&    -0.78$\pm$0.31&    2.11\\
Bo044 &    8.25$\pm$2.26&    -1.14$\pm$0.37&    1.51\\
Bo051 &   11.25$\pm$2.51&    -1.00$\pm$0.13&    1.14\\
Bo054 &    1.28$\pm$0.03&    -0.45$\pm$0.17&    3.00\\
Bo056 &    3.50$\pm$0.78&    -0.06$\pm$0.10&    3.54\\
Bo057 &    6.75$\pm$1.95&    -2.12$\pm$0.32&    1.23\\
Bo059 &   20.00$\pm$1.01&    -1.36$\pm$0.52&    1.83\\
Bo061 &    7.25$\pm$0.87&    -0.73$\pm$0.28&    1.73\\
Bo063 &    9.75$\pm$2.58&    -0.87$\pm$0.33&    1.43\\
Bo064 &   18.00$\pm$2.47&    -1.55$\pm$0.30&    1.85\\
Bo068 &    1.80$\pm$0.04&    -0.29$\pm$0.59&    1.33\\
Bo073 &    7.25$\pm$0.87&    -0.64$\pm$0.46&    1.78\\
Bo076 &    1.68$\pm$0.10&    -0.72$\pm$0.06&    1.78\\
Bo082 &    6.00$\pm$0.87&    -0.80$\pm$0.18&    3.35\\
Bo086 &   15.25$\pm$2.72&    -1.74$\pm$0.17&    1.25\\
Bo088 &    7.75$\pm$0.84&    -1.81$\pm$0.06&    0.60\\
Bo093 &   20.00$\pm$5.44&    -1.03$\pm$0.12&    1.64\\
Bo094 &    3.25$\pm$0.30&    -0.17$\pm$0.45&    1.34\\
Bo096 &    9.00$\pm$1.65&    -0.26$\pm$0.43&    1.81\\
Bo097 &    2.30$\pm$0.23&    -1.21$\pm$0.13&    1.19\\
Bo098 &    3.00$\pm$0.11&    -0.67$\pm$0.58&    1.41\\
Bo102 &    1.80$\pm$0.24&    -1.57$\pm$0.10&    2.05\\
Bo103 &    7.75$\pm$1.15&    -0.56$\pm$0.62&    1.17\\
Bo105 &   10.25$\pm$1.82&    -1.13$\pm$0.32&    0.56\\
Bo106 &   20.00$\pm$7.11&    -0.86$\pm$0.68&    2.18\\
Bo107 &    5.25$\pm$1.84&    -1.18$\pm$0.30&    2.25\\
Bo110 &    8.00$\pm$2.43&    -1.06$\pm$0.12&    2.00\\
Bo116 &    7.50$\pm$0.76&    -0.88$\pm$0.12&    0.67\\
Bo125 &    4.00$\pm$0.71&    -1.52$\pm$0.08&    1.62\\
Bo127 &    9.75$\pm$3.03&    -0.80$\pm$0.14&    3.88\\
Bo130 &    6.50$\pm$1.74&    -1.28$\pm$0.19&    1.71\\
Bo134 &    5.50$\pm$2.28&    -0.64$\pm$0.08&    5.47\\
Bo135 &   20.00$\pm$2.10&    -1.62$\pm$0.04&    1.18\\
Bo137 &    5.50$\pm$1.08&    -1.21$\pm$0.29&    1.39\\
Bo143 &    4.75$\pm$0.76&     0.09$\pm$0.42&    1.99\\
Bo148 &    7.50$\pm$2.34&    -1.15$\pm$0.34&    3.07\\
Bo149 &    4.25$\pm$1.11&    -1.35$\pm$0.25&    3.38\\
Bo151 &   12.50$\pm$2.61&    -0.75$\pm$0.18&    2.05\\
Bo152 &   17.50$\pm$4.76&    -0.87$\pm$0.49&    2.12\\
Bo153 &    8.00$\pm$0.83&    -0.08$\pm$0.33&    1.10\\
Bo154 &    1.61$\pm$0.17&    -0.45$\pm$0.63&    4.36\\
Bo158 &   20.00$\pm$5.57&    -1.02$\pm$0.02&    3.06\\
Bo161 &   20.00$\pm$1.00&    -1.60$\pm$0.48&    3.64\\
Bo163 &   11.75$\pm$1.60&    -0.36$\pm$0.27&    1.87\\
Bo164 &    1.61$\pm$0.18&    -0.09$\pm$0.40&    1.88\\
Bo165 &   11.50$\pm$3.08&    -1.80$\pm$0.32&    1.49\\

\hline
\end{tabular}
\end{center}
\end{table}

\begin{table}
\caption{continued}
\begin{center}
\begin{tabular}{cccc}
\hline  ID&   Age &  Metallicity &   $\chi_{\rm min}^2$ \\
          & (Gyr) &    [Fe/H]    & (per degree of freedom) \\
\hline
Bo167 &    5.50$\pm$1.12&    -0.42$\pm$0.23&    1.15\\
Bo171 &   10.75$\pm$1.87&    -0.41$\pm$0.04&    2.49\\
Bo174 &   18.50$\pm$1.88&    -1.67$\pm$0.27&    1.27\\
Bo178 &   15.50$\pm$2.37&    -1.51$\pm$0.12&    1.03\\
Bo179 &    9.75$\pm$2.17&    -1.10$\pm$0.02&    0.97\\
Bo180 &    6.25$\pm$1.37&    -1.19$\pm$0.07&    1.67\\
Bo182 &   16.75$\pm$2.93&    -1.24$\pm$0.12&    1.33\\
Bo183 &    3.00$\pm$0.09&    -0.19$\pm$0.31&    1.05\\
Bo184 &    1.68$\pm$0.08&    -0.37$\pm$0.40&    1.63\\
Bo185 &    9.75$\pm$3.24&    -0.76$\pm$0.08&    2.67\\
Bo190 &   12.75$\pm$2.24&    -1.03$\pm$0.09&    0.57\\
Bo193 &   15.50$\pm$3.92&    -0.44$\pm$0.17&    1.49\\
Bo197 &    0.25$\pm$0.03&    -0.43$\pm$0.36&    2.67\\
Bo201 &   10.75$\pm$4.26&    -1.06$\pm$0.21&    2.64\\
Bo203 &    6.25$\pm$0.86&    -0.90$\pm$0.32&    1.45\\
Bo204 &    7.00$\pm$0.93&    -0.80$\pm$0.17&    2.02\\
Bo205 &   20.00$\pm$2.29&    -1.34$\pm$0.13&    1.85\\
Bo206 &   20.00$\pm$2.50&    -1.45$\pm$0.10&    3.30\\
Bo209 &   20.00$\pm$0.74&    -1.37$\pm$0.13&    1.78\\
Bo211 &   19.00$\pm$1.68&    -1.67$\pm$0.52&    1.03\\
Bo213 &    7.50$\pm$1.61&    -1.02$\pm$0.11&    1.89\\
Bo214 &    4.00$\pm$1.50&    -1.00$\pm$0.61&    2.63\\
Bo217 &   10.25$\pm$2.15&    -0.93$\pm$0.14&    1.05\\
Bo218 &   19.00$\pm$2.30&    -1.19$\pm$0.07&    2.06\\
Bo220 &    6.00$\pm$0.83&    -1.21$\pm$0.09&    0.96\\
Bo221 &   16.75$\pm$3.57&    -1.29$\pm$0.04&    1.47\\
Bo222 &    7.75$\pm$1.46&    -0.93$\pm$0.95&    5.09\\
Bo224 &    5.50$\pm$1.00&    -1.80$\pm$0.05&    0.45\\
Bo225 &   15.50$\pm$4.89&    -0.67$\pm$0.12&    2.52\\
Bo228 &   10.00$\pm$1.86&    -0.65$\pm$0.66&    0.98\\
Bo235 &    4.00$\pm$0.59&    -0.72$\pm$0.26&    1.65\\

\hline
Bo037$^a$ &12.4$\pm$3.2 &  $-$1.07$\pm$0.20&    1.43\\
\hline
\end{tabular}
\end{center}
{Note: $^a$ values taken from \citet{Ma06b}.}
\end{table}

\begin{table}
\caption{Check of the effects of the photometric uncertainties.}
\label{t2.tab}
\begin{center}
\begin{tabular}{cccc}
\hline
Name&   &  Best-fitting  age (Gyr) &  \\
\cline{2-4}
         & (original errors) & ($0.5 \times$ original) & ($2\times$ original) \\
\hline
Bo015 &   1.43$\pm$0.04   &    1.43$\pm$0.04     &    1.43$\pm$0.05     \\
Bo137 &   5.50$\pm$1.08   &    5.75$\pm$1.09     &    4.75$\pm$1.09     \\
Bo228 &  10.00$\pm$1.86   &   10.50$\pm$1.72     &    9.75$\pm$2.11     \\
Bo178 &  15.50$\pm$2.37   &   16.75$\pm$2.36     &   14.50$\pm$2.32     \\
Bo135 &  20.00$\pm$2.10   &   20.00$\pm$1.34     &   18.50$\pm$1.89     \\
\hline
\end{tabular}
\end{center}
\end{table}

\subsection{Check of the importance of the photometric uncertainties}
\label{uncertainties.sec}

\begin{figure}
\resizebox{\hsize}{!}{\rotatebox{-90}{\includegraphics{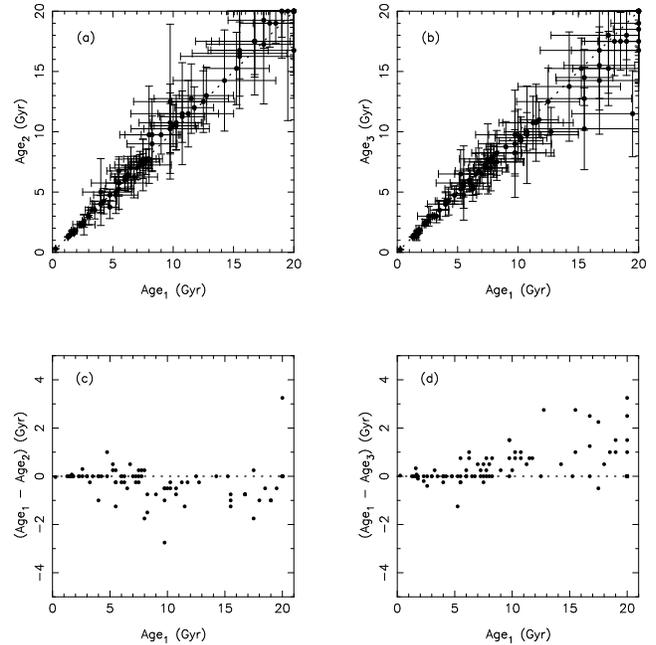}}}
\caption{Comparison of the best-fitting GC ages and their
dependence on the photometric uncertainties assumed. Subscripts 1,
2 and 3 refer to the assumption of, respectively, the actual
photometric uncertainties, half and twice these uncertainties. We
have omitted the error bars in panels c and d for reasons of
clarity; the dotted lines indicate the loci of equality.}
\label{fig:unc}
\end{figure}

We will explore and quantify to what extent the photometric
uncertainties affect our results. To this end, we randomly selected a
number of GCs, while making sure that they spanned our entire age
range. The results for a small subsample of our GCs are listed in
Table 3. From the left to the right column on the same line, the
photometric uncertainties vary from their original values (1), half
the original (2), and twice the original (3). From this initial
comparison it is clear that the photometric uncertainties adopted do
not affect our results significantly, in view of the large
uncertainties in the derived ages, but we note small differences in
the best-fitting ages towards the high-age end. In Fig. \ref{fig:unc}
we show the results for the entire sample. The subscripts, 1, 2 and 3,
refer to conditions (1), (2) and (3) above, respectively. A comparison
of Figs. \ref{fig:unc}a and c on the one hand, and b and d on the
other show that (i) the best-fitting age estimates are essentially
independent of the photometric uncertainty assumed for ages up to
about 8 Gyr; (ii) for older ages, the best-fitting ages tend to be
overestimated by $\la 1.5$ Gyr, in the majority of cases, if the
photometric uncertainties are halved, and (iii) underestimated by $\la
2$ Gyr, for most objects, if the photometric uncertainties are
doubled. We also note, however, that the actual uncertainties on the
derived ages are well in excess of these small systematic effects at
old age; for this reason (and for improved clarity) we have omitted
the error bars in Figs. \ref{fig:unc}c and d. Nevertheless, we will
need to keep these small systematic offsets in mind when assessing the
absolute uncertainties in our best-fitting age estimates, which we
will do in the next section.

\subsection{Comparison of age determinations}
\label{agecomp.sec}

\begin{figure}
\resizebox{\hsize}{!}{\rotatebox{-90}{\includegraphics{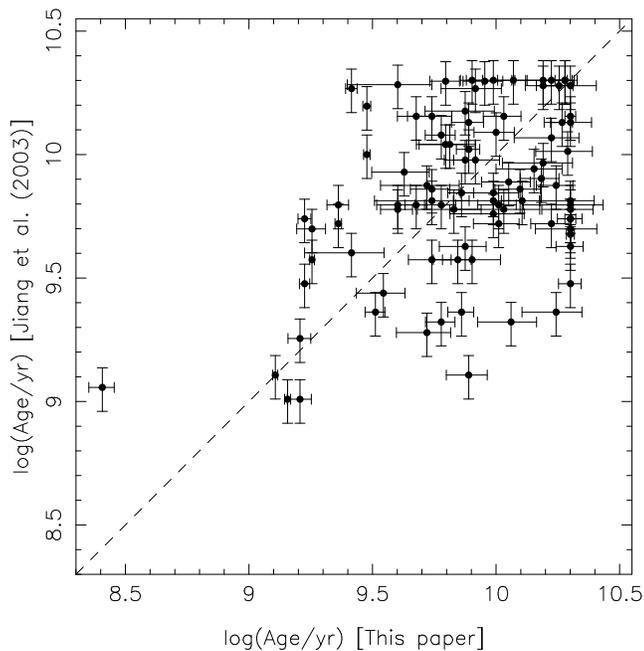}}}
\vspace{0.2cm} \caption{Comparison of age determinations for the
M31 GCs in common between \citet{jiang03} -- based on \citet{bc96}
and on the BATC multi-band photometry; the error bars correspond
to their ``typical'' 20 per cent uncertainties -- and this paper.}
\label{fig5}
\end{figure}

\citet{jiang03} estimated the ages of their 172 M31 GCs, based on
only the BATC data and on the SSP models of BC96; the reddening
values adopted by \citet{jiang03} were also from \citet{bh00}.
However, \citet{jiang03} only used the BC96 models for three
metallicities, i.e., 0.0004, 0.004 and 0.02, and did not linearly
interpolate to find the best-fitting metallicities. For old GCs,
the age/metallicity degeneracy becomes important, so it is
reasonable to postulate that \citet{jiang03} did not estimate the
ages of their GC sample as accurately as we have done in this
paper. Figure \ref{fig5} shows a comparison of the age
determination for the M31 GCs in common between \citet{jiang03}
and this paper. It is immediately clear that there is a
significant difference between \citet{jiang03} and our results.

\begin{figure}
\resizebox{\hsize}{!}{\rotatebox{-90}{\includegraphics{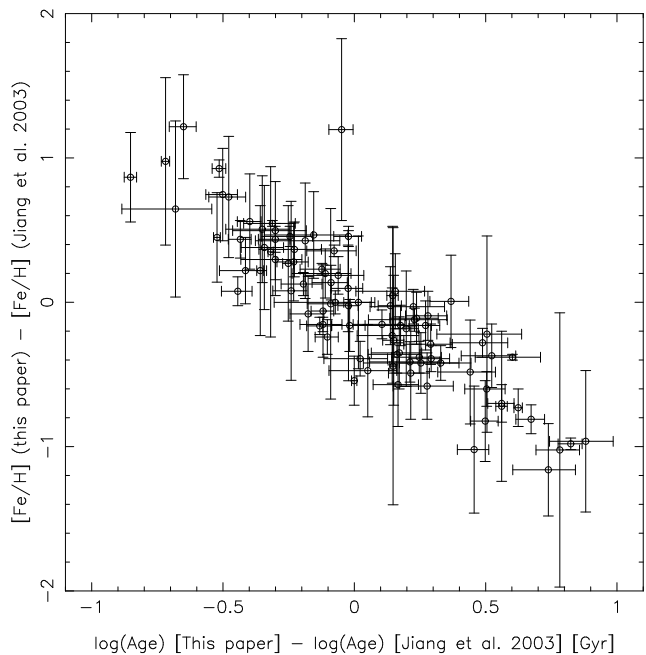}}}
\caption{Age and metallicity differences between the present paper and
the determinations by \citet{jiang03}; see text for details.}
\label{degeneracy.fig}
\end{figure}

In order to have confidence in our age determinations in this paper,
we must make an effort to understand the main cause of these
significant differences between the age estimates presented here and
those of \citet{jiang03}. Since we suspected that the age-metallicity
degeneracy may be the principal culprit in this regard, we used Table
3 in \citet{jiang03} to produce Fig.  \ref{degeneracy.fig}. Here, we
show the difference in age versus the difference in metallicity for
the M31 GCs in common between our samples; we used a straightforward
conversion from $Z$, as listed in \citet{jiang03}, and [Fe/H],
assuming for the sake of simplicity that these parameters are one and
the same. The error bars along the age axis are a combination of the
uncertainties in age from both the present paper and the 20 per cent
uncertainties quoted in \citet{jiang03}; the error bars along the
metallicity axis represent the uncertainties in [Fe/H] from Table 2
only. From Fig. \ref{degeneracy.fig} it is immediately obvious that
there is indeed a very strong age-metallicity degeneracy present among
our GC sample; this is also (and just as significantly so) the case if
we had used ``Age'' instead of ``log(Age)'' as our horizontal axis
(not shown). The correlation shown in this figure is excellent, and in
the sense expected if our new age estimates are indeed more
accurate\footnote{We note that the tight correlation is predominantly
driven by the age-metallicity degeneracy evident in the
\citet{jiang03} results: the distribution of data points in the
$\Delta \log({\rm Age})$ versus [Fe/H] \citep{jiang03} plane shows a
clear general trend of decreasing metallicity with increasingly large
ages in the \citet{jiang03} results; such a clear correlation is not
evident in the $\Delta \log({\rm Age})$ versus [Fe/H] plane defined by
our new data.}; the scatter is likely partially owing to the discrete
steps in metallicity adopted by \citet{jiang03}, as well as to the
intrinsic uncertainties in the metallicity determination and
possibly to a residual (but small; see footnote 4) age-metallicity
degeneracy in the age determinations presented here. In any case, the
scatter in the relationship is well inside the $\sim 1 \sigma$
level. Thus, here we have shown that we fully understand the
significant differences between our age determinations and those of
\citet{jiang03}. We believe to have presented significant improvements
in the present paper, as well as clear evidence that the previous age
determinations of statistically large samples of M31 GCs were
more significantly affected by the age-metallicity degeneracy
affecting old GCs than our new age estimates.

\begin{figure}
\resizebox{\hsize}{!}{\rotatebox{-90}{\includegraphics{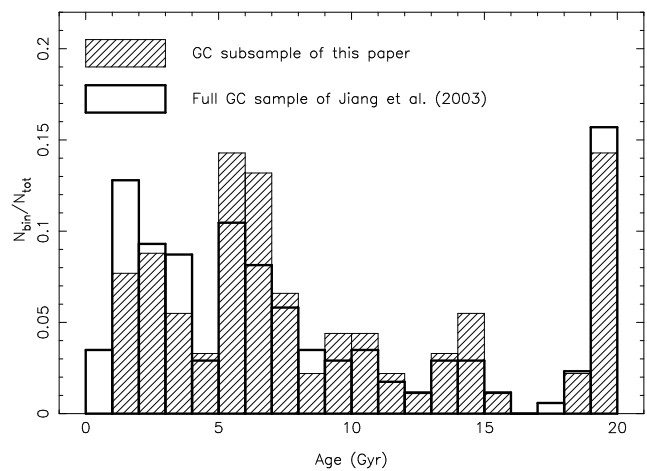}}}
\vspace{0.2cm}
\caption{Age distributions of the M31 GCs in \citet{jiang03} and in
this paper, both based on the ages determined by \citet{jiang03}. The
age distribution of the GCs discussed in this paper can therefore
immediately be compared to the full distribution of the
\citet{jiang03} sample.} \label{fig7}
\end{figure}

We will now explore whether our GC sample constitutes a random
subsample of the \citet{jiang03} GC sample. Fig. \ref{fig7} shows the
age distribution for the full \citet{jiang03} GC sample, as well as
for the full GC sample selected in this paper. Both distributions are
based on the age determinations of \citet{jiang03}. In
Fig. \ref{fig7}, the numbers have been normalized, i.e., in each bin
the number of objects is divided by the total number of objects. To
first order, Fig. \ref{fig7} shows that our sample constitutes a
randomly selected subsample of the \citet{jiang03} GC sample; the main
features of the age distribution are apparent in both
distributions. Any apparent differences, such as the relative ratio of
the two age peaks at ages below 10 Gyr, can be attributed to
Poissonian uncertainties at the $\sim 1-2 \sigma$ level. To quantify
the similarities and differences between both age distributions, we
employed a Kolmogorov-Smirnov (KS) test. The maximum value of the
absolute difference is $D_{\rm max}=0.1259$ for these two samples,
containing 91 and 172 data points, respectively. The probability of
obtaining this value for $D_{\rm max}$ is 28.3 per cent, which means
that there is a small, 28.3 per cent probability that the two
distributions are different.

\subsection{The M31 GC age distribution}

\begin{figure}
\resizebox{\hsize}{!}{\rotatebox{-90}{\includegraphics{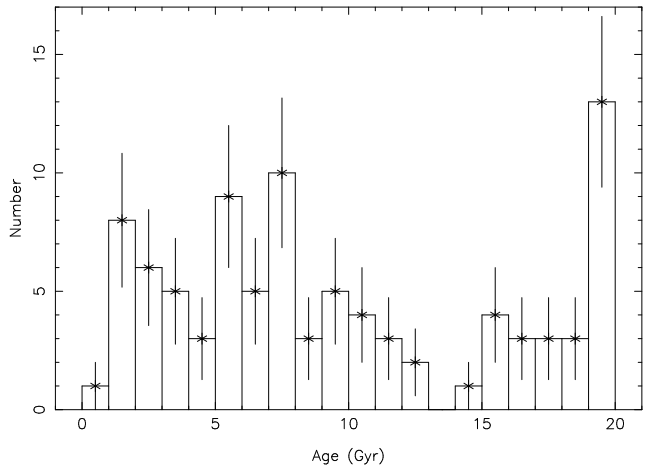}}}
\vspace{0.3cm} \caption{Age distribution of our 91 sample GCs.
Poissonian statistical uncertainties are shown.} \label{fig8}
\end{figure}

Fig. \ref{fig8} shows the (newly determined) age distribution of our
sample GCs. It shows that there exist young and intermediate-age
subpopulations among the M31 GC population, peaking at ages of $\sim
3$ and $\sim 8$ Gyr, respectively. The ``usual'' complement of GCs of
similar age as the old Galactic GCs, previously discussed by
\citet{bh00}, \citet{bb04}, \citet{bur04}, and \citet{puz05}, is -- of
course -- present as well. Assuming Poissonian statistical errors, the
two (young and intermediate-age) subpopulations are indeed significant
at the $\sim 3 \sigma$ level. In order to assess the robustness of
these features in the newly derived age distribution of our M31 GC
sample, we need to take into account the corresponding uncertainties
in age, as well as the bin size (in Gyr) used to construct the age
distribution.

A comparison of the uncertainties in our derived GC ages (obtained
from our SED fitting routine) as a function of age (not shown),
reveals that the errors increase significantly with increasing
age. This is simply owing to the reduced age resolution of older
broad-band SEDs, since SSP colours are not as sensitive to age
variations at old ages (i.e., in excess of a few Gyr) compared to
their earlier evolution. For GC ages up to $\sim 5$ Gyr, the
uncertainties in our derived cluster ages are generally $\la 1$ Gyr,
in most cases significantly so. Both the spread and the upper limit of
the age uncertainties increase, the upper limit roughly linearly. The
latter reaches $\sim 1.5$ Gyr at an age of $\sim 7-8$ Gyr. This is, in
fact, the age range of interest for our assessment of the robustness
of the doubly-peaked M31 GC age distribution. Fortunately, as we
discussed in Section \ref{uncertainties.sec}, the derived ages are
insensitive to the photometric uncertainties assumed at these young
and intermediate ages. We therefore adopt the uncertainties on the
derived ages as the basis for our bin sizes; the doubly-peaked age
distribution remains a significant feature at the $\sim 3\sigma$ level
for age bins of up to 1.5 Gyr. We conclude, therefore, that this
feature is real and robustly detected. In addition, in Fig. 9 we
plot the relative uncertainties in our derived GC ages as a function
of age. It is clear that, although the absolute age errors increase
significantly with increasing age, the relative uncertainties are
rather similar over the full range of derived ages, with a mean of
$\Delta {\rm Age / Age} \simeq 0.17 \pm 0.10$ (1$\sigma$ spread).

The age peak at $\sim 7-8$ Gyr appears to roughly coincide with the
intermediate-age GC population aged between $\sim 5$ and 8 Gyr
discussed by \citet{puz05}, which they found to have a mean
metallicity of [Z/H] $\approx -0.6$ dex (although they also note that
their sample is biased towards high-metallicity objects). It is worth
noting that a similar intermediate-age star-forming event may have
affected the resolved stellar population, at least along M31's minor
axis: \citet{brown03} found indications of an intermediate-age
resolved stellar population, of $\sim 6-8$ Gyr and with [Fe/H] $>
-0.5$ dex, from their {\sl HST}/ACS observations.

\begin{figure}
\resizebox{\hsize}{!}{\rotatebox{-90}{\includegraphics{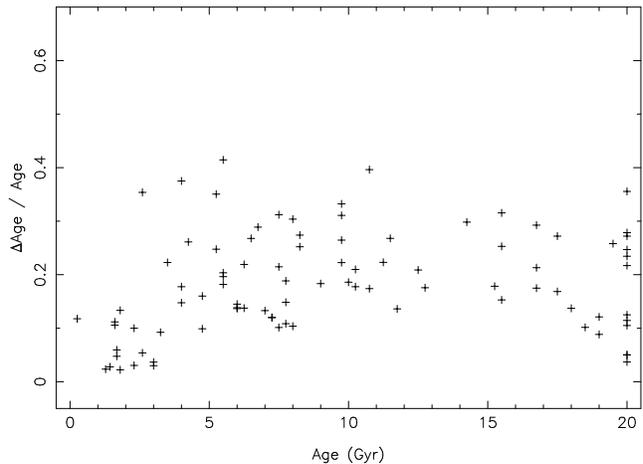}}}
\vspace{0.3cm}
\caption{Relative uncertainties in the M31 GC ages determined in this
paper, as a function of age.} \label{fig9}
\end{figure}

The young-age peak at $\sim 3$ Gyr we detect in this paper has not
been discussed before; in view of the small age uncertainties (see
above) at young ages, it is unlikely that this population represents
the $\la 2$ Gyr-old population of BLCCs of \citet{fp05}. Instead, we
argue that there may have been an additional violent star-forming
event that triggered the formation of a GC subpopulation in the disc
of M31 some 3 Gyr ago.

It is well-known that the abundance distributions of GCs in many
galaxies are bimodal, including that of the M31 GC system
\citep[e.g.][]{bh00,per02}. \citet{forbes97} found that the
metal-rich GCs in elliptical and cD galaxies are closely coupled
to the resolved stellar populations of their parent galaxies, but
the metal-poor GCs are largely independent of the galaxies. They
concluded that the metal-poor GCs were formed at the time of
galaxy formation, and the metal-rich GCs at a later stage. If so,
this may imply that the age distribution of GCs should be bi- or
multi-modal. Despite our incomplete coverage of the M31 halo and
the extreme outer disc regions along the galaxy's major axis,
based on Fig. \ref{fig8} we can confirm that, in general, the age
distribution of the M31 GCs is unlikely to be best described by a
monomodal distribution \citep[see also][for an extensive
discussion, based on their high-metallicity biased GC
sample]{puz05}.

The age distributions of both the full M31 GC sample and of the field
stars, combined with the detection of a metal-poor GC population
obeying thin-disc kinematics \citep{mor04}, poses serious problems for
our understanding of the formation and evolution of the galaxy's
disc. On the one hand, the tightly constrained kinematics argue for a
relatively quiescent thin-disc evolution since early times \citep[but
see][]{abadi}, while on the other hand the formation of massive star
clusters requires violent conditions (such as galaxy mergers) rather
than {\it in situ} formation \citep[see][for a
discussion]{fp05,puz05}.

\subsection{A Relationship between Age and Metallicity?}

\begin{figure}
\resizebox{\hsize}{!}{\rotatebox{-90}{\includegraphics{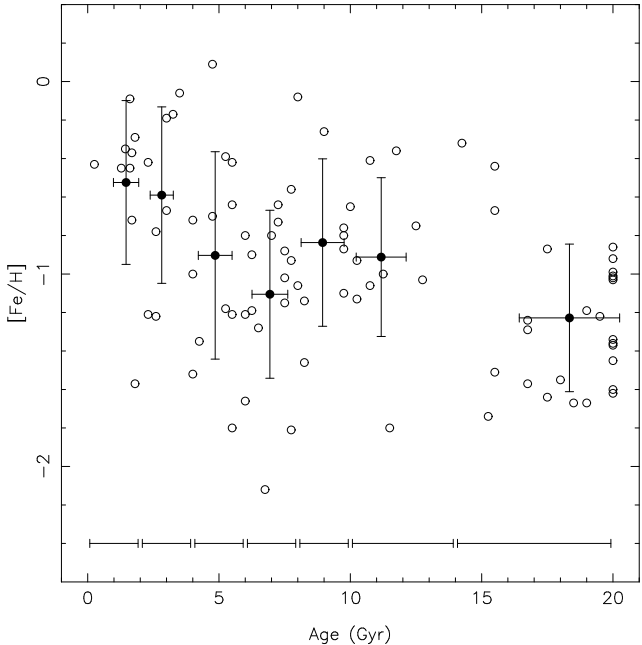}}}
\vspace{0.1cm} \caption{Distribution of the ages and metallicities
of our 91 M31 GCs. Error bars have been omitted for reasons of
clarity. The filled circles with their error bars represent the
mean distribution and the Gaussian $\sigma$'s of the data points;
the age ranges used to obtain the mean values are indicated at the
bottom of the figure by horizontal bars.} \label{fig10}
\end{figure}

The Galactic GC system is well-known to obey a clear age-metallicity
relationship, in the sense of older GCs being characterised by
generally lower metallicities \citep[e.g.][]{sc89,chaboyer96}.
\citet{Ma06b} simultaneously obtained ages and metallicities for 33
M31 GCs and GC candidates, based on the BC03 SSP models, but did not
find any significant relationship between their ages and
metallicities. However, because of the age-metallicity degeneracy
affecting the analysis done to the \citet{Ma06b} GC sample, the ages
and metallicities could not be obtained sufficiently accurately.
Therefore, using the ages and metallicities of the M31 GCs discussed
in this paper, we re-analysed the M31 GC sample. Fig. \ref{fig10}
shows the metallicity as a function of age for the GC sample discussed
in this paper. It appears that a general trend between GC age and
metallicity may exist, in the sense that the metal-poorer clusters are
older, although the scatter in metallicity at a given age is
significant. It may be better to postulate instead that there is a
general lack of both low-metallicity younger GCs and of the
highest-metallicity clusters at the oldest ages. Error bars have been
omitted for reasons of clarity.

In order to assess the robustness of these statements, we produce the
run of the mean of the data points across the age range, shown by the
filled circles with the error bars. The error bars indicate the
Gaussian $\sigma$'s of the data points in each age range, indicated by
the horizontal bars at the bottom of the figure.  Table 4 is a
compilation of the data points obtained for the mean values shown in
Fig. \ref{fig10}. The trend outlined by these mean data points
confirms the statements made here regarding the possible existence of
an age-metallicity relation among the M31 GC sample. In view of our
carefully validated random sampling of our M31 GCs, we believe that
this age-metallicity relation is representative of the entire M31 GC
population, providing that our positional (field of view) restrictions
did not invalidate the random sampling (see Section 2.1). In view of
the latter concern, we point out that our GC sample may be affected by
an underrepresentation of BLCCs; these objects are mostly aged between
$\sim 500$ Myr and $\sim 2$ Gyr, and may in fact have metallicities in
the range $-1.0 \la {\rm [Fe/H]} \la 0.0$ \citep{fp05}. If BLCCs are
indeed underrepresented in our sample, their additional inclusion
would, in fact, strengthen the age-metallicity relationship revealed
in this paper.

A similar, although somewhat less robust result was found by
\citet{jiang03}, who reported to have determined that the more
metal-poor M31 GCs appear to be older than their more metal-rich
counterparts. This is, in essence, a similar conclusion as reached by
\citet{bh00apjl} and \citet{puz05}, while the combination of
positional information and the age-metallicity relationship found here
is supportive of the existence of a small radial metallicity gradient
in M31, discussed by \citet{bh00,per02,puz05} and references therein.

In Section 4.2, we estimated the ages for our 91 GCs using a
$\chi^2$ minimisation method, in which we included the (systematic)
model uncertainties following \citet{charlot96}. We added these in
quadrature to the observational uncertainties, in a similar fashion as
\citet{wu05} and \citet{Ma06b}. However, we should be aware that this
procedure could potentially somewhat confuse our statistical results,
since the model uncertainties might introduce biases in the
age-metallicity correlation, while the observational uncertainties
smooth and puff-up such correlations. On the other hand, we point out
that the model uncertainties are of the same order as the measurement
uncertainties quoted by \citet{jiang03}, at least for a large fraction
of their measurements, so that the effect of adding the model
uncertainties {\it in quadrature} implies, in essence, less than a
doubling of the uncertainties, similar to what we did in Section
4.3. As such, we are confident that we have not introduced major
biases into our fitting routine by taking into account the
uncertainties inherent in the models (note that we only increase the
uncertainties following this procedure; we did not add a systematic
offset to the measurements.

\begin{table}
\caption{Mean age-metallicity relation for the M31 GCs.}
\label{t3.tab}
\begin{center}
\begin{tabular}{ccccc}
\hline
Age range & Mean age & $\sigma$ & Mean [Fe/H] & $\sigma$ \\
(Gyr)     & (Gyr)    & (Gyr)    & (dex)       & (dex)    \\
\hline
$ 0 \le {\rm Age} <  2$ &  1.46 & 0.48 & $-$0.52 & 0.43 \\
$ 2 \le {\rm Age} <  4$ &  2.82 & 0.44 & $-$0.59 & 0.46 \\
$ 4 \le {\rm Age} <  6$ &  4.85 & 0.64 & $-$0.90 & 0.54 \\
$ 6 \le {\rm Age} <  8$ &  6.94 & 0.68 & $-$1.11 & 0.44 \\
$ 8 \le {\rm Age} < 10$ &  8.94 & 0.82 & $-$0.84 & 0.43 \\
$10 \le {\rm Age} < 14$ & 11.18 & 0.95 & $-$0.91 & 0.41 \\
$14 \le{\rm Age}\le 20$ & 18.34 & 1.91 & $-$1.23 & 0.38 \\
\hline
\end{tabular}
\end{center}
\end{table}

\section{Summary and Conclusions}
\label{Conclusions.sec}

In this paper, we accurately re-determined the ages of 91 M31 GCs,
based on improved data, updated theoretical stellar synthesis models
and sophisticated fitting methods. In particular, we used photometric
measurements of the 2MASS, which can partially break the
age-metallicity degeneracy, in combination with optical photometry. We
showed robustly that previous age determinations based on photometric
data were affected significantly by this age-metallicity degeneracy.

Except for one cluster, the ages of our other sample GCs are all older
than 1 Gyr. Their age distribution shows populations of young and
intermediate-age GCs, peaking at $\sim 3$ Gyr and $\sim 8$ Gyr
respectively, as well as the ``usual'' complement of well-known old
GCs, of similar age as the majority of the Galactic GCs. The young-age
peak at $\sim 3$ Gyr we detect in this paper has not been discussed
before; in view of the small age uncertainties at young ages, it is
unlikely that this population represents the $\la 2$ Gyr-old
population of BLCCs of \citet{fp05}. Instead, we argue that there may
have been an additional violent star-forming event that triggered the
formation of a GC subpopulation in the disc of M31 some 3 Gyr ago.

The distributions of the ages of both the GCs and the field stars,
combined with the existence of a metal-poor GC population in M31
obeying thin-disc kinematics, poses serious problems for our
understanding of the formation and evolution of the galaxy's disc.
While the tightly constrained kinematics argue for a relatively
quiescent thin-disc evolution since early times, the formation of
massive star clusters requires violent conditions (such as galaxy
mergers) rather than {\it in situ} formation.

Our results also show that although there is significant scatter in
metallicity at any age, there is a noticeable lack of young metal-poor
and old metal-rich GCs, which might be indicative of an underlying
age-metallicity relationship among the M31 GC population.

\section*{Acknowledgments}
We thank Uta Fritze and Simon Goodwin for useful comments on aspects
of this paper; we are also indebted to the referee for his/her
thoughtful comments and insightful suggestions that improved this
paper greatly. This work has been supported by the Chinese National
Key Basic Research Science Foundation (NKBRSF TG199075402) and by the
Chinese National Natural Science Foundation, No. 10473012 and
10333060.

\end{document}